\documentclass{aa}  

\usepackage{epsfig}

\usepackage{graphics}

\usepackage{subfigure}

\usepackage{natbib}

\usepackage{amsmath}

\usepackage{amssymb}

\usepackage{upgreek}

\usepackage{enumerate}

\bibpunct{(}{)}{,}{a}{}{,} 

\newcommand{\frb}{FRB~121102}

\begin{document}

\title{A model for the repeating FRB~121102 in the AGN scenario}

\author{F.~L.~Vieyro\inst{1,2}, G.~E.~Romero\inst{2,3}, V.~Bosch-Ramon\inst{1}, B.~Marcote\inst{4}, \and M.~V.~del Valle\inst{5}}

\institute{Departament de F\'{i}sica Qu\`antica i Astrof\'{i}sica, Institut de Ci\`encies del Cosmos (ICCUB), Universitat de Barcelona, IEEC-UB, Mart\'{i} i Franqu\`es 1, E08028 Barcelona, Spain
\and Instituto Argentino de Radioastronom\'{\i}a (IAR, CCT La Plata, CONICET; CICPBA), C.C.5, (1984) Villa Elisa, Buenos Aires, Argentina
\and Facultad de Ciencias Astron\'omicas y Geof\'{\i}sicas, Universidad Nacional de La Plata, Paseo del Bosque s/n, 1900, La Plata, Argentina
\and Joint Institute for VLBI ERIC, Postbus 2, 7990 AA Dwingeloo, The Netherlands
\and Institute of Physics and Astronomy, University of Potsdam, 14476 Potsdam-Golm, Germany}

\offprints{F. L. Vieyro \\ \email{fvieyro@fqa.ub.edu}}

\titlerunning{Model of a repeating fast radio burst}

\authorrunning{Vieyro, Romero, Bosch-Ramon, et al.}

\abstract
{Fast radio bursts, or FRBs, are transient sources of unknown origin. Recent radio and optical observations have provided strong evidence for an extragalactic origin of the phenomenon and the precise localization of the repeating FRB~121102. Observations using the Karl G. Jansky Very Large Array (VLA) and very-long-baseline interferometry (VLBI) have revealed the existence of a continuum non-thermal radio source consistent with the location of the bursts in a dwarf galaxy. All these new data rule out several  models that were previously proposed, and impose stringent constraints to new models.}
{We aim to model FRB~121102 in light of the new observational results in the active galactic nucleus (AGN) scenario.}
{We propose a model for repeating FRBs in which a non-steady relativistic $e^\pm$-beam, accelerated by an impulsive magnetohydrodynamic (MHD)-driven mechanism, interacts with a cloud at the centre of a star-forming dwarf galaxy. The interaction generates regions of high electrostatic field called cavitons in the plasma cloud. Turbulence is also produced in the beam. These processes, plus particle isotropization, the interaction scale, and light retardation effects, provide the necessary ingredients for short-lived, bright coherent radiation bursts.}
{The mechanism studied in this work explains the general properties of FRB~121102, and may also be applied to other repetitive FRBs.}
{Coherent emission from electrons and positrons accelerated in cavitons provides a plausible explanation of FRBs.}


\maketitle

\section{Introduction}

Fast radio bursts (FRBs) are bright transient flashes of cosmic origin with durations of a few milliseconds detected at radio wavelengths. They were discovered by \citet{lorimer2007} and found to exhibit large dispersion measures (DM).  These dispersions are in excess of the contribution expected from the electron distribution of our Galaxy, hence suggesting an extragalactic, cosmological origin. Most of the eighteen known bursts have been detected so far with the Parkes radio telescope \citep[e.g.][]{thornton2013,petroff2016}.  Only a couple of them were found with the Arecibo and Green Bank telescopes \citep{spitler2014, spitler2016, masui2015}.

The physical origin of FRBs remains a mystery. The putative extragalactic distances and the extremely rapid variability imply brightness temperatures largely beyond the Compton limit for incoherent synchrotron radiation \citep[e.g.][]{katz2014}. Thus a coherent origin of the radiation seems certain. Models proposed so far can be divided into those of catastrophic nature, in which the source does not survive the production of the burst, and those that can repeat. A non-unique FRB population is possible, with different types of sources, as is the case with the gamma-ray bursts \citep[see][for a review]{katz2016}.

The unambiguous identification of the counterparts of FRBs at other wavelengths is a very difficult task because of the extremely short life span of the events at radio frequencies, their appearance from random directions in the sky, and the large uncertainties in the determination of their precise positions. A huge step towards the clarification of the origin and nature of these happenings was the recent direct localization of an FRB  and its host by \citet{chatterjee2017}. These authors achieved the sub-arc second localization of FRB~121102, the only known repeating FRB, using high-time-resolution radio interferometric observations that directly imaged the bursts. They found that FRB~121102 originates very close to a faint and persistent radio source with a continuum spectrum consistent with non-thermal emission and a faint optical counterpart. This latter optical source has been identified by \citet{tendulkar2017} as a low-metallicity, star-forming, dwarf galaxy at a redshift of $z = 0.19273(8)$, corresponding to a luminosity distance of 972~Mpc. Further insights on the persistent radio source were provided by \citet{marcote2017} through very-long-baseline radio interferometric observations. Marcote et al. were able to simultaneously detect and localize both, the bursts and the persistent radio source, on milliarcsecond scales. The bursts are found to be consistent with the location of the persistent radio source within a projected linear separation of less than $40$~pc, $12$~mas angular separation, at 95\% confidence, and thus both are likely related. The unambiguous association of FRB~121102 with persistent radio and optical counterparts, along with the identification of its host galaxy, impose, for the first time, very strict constraints upon theoretical models for FRBs, beyond the general limits imposed by variability timescales and energy budgets.

\citet{romero2016} show that under certain conditions, a turbulent plasma hit by a relativistic jet can emit short bursts consistent with the ones observed in FRBs. In this paper we apply the latter model to FRB~121102, based on the idea that the multiple observed bursts are the result of coherent phenomena excited in turbulent plasma by the interaction of a sporadic relativistic $e^\pm$-beam or jet, which originates from a putative somewhat massive black hole in the central region of the observed dwarf galaxy and ambient material. Our model can account for the different properties known so far for FRB~121102 and can be tested through observations of other repetitive FRBs. In what follows we first detail the known features of FRB~121102 and its host that are relevant for the involved physics (Sect. \ref{sect2}), then we describe our model (Sect. \ref{sect3}) and its application to FRB~121102 (Sect. \ref{sect4}), and we finally offer some discussion and our conclusions (Sects. \ref{sect.5} and \ref{sect.6}).

\section{Main facts about FRB~121102}\label{sect2}

Fast radio burst~121102 is the only known source of its class that presents repeated bursts with consistent DM and sky localization \citep{spitler2014,spitler2016,scholz2016}. This implies that the source is not annihilated by the production mechanism of the bursts. Most of the individual bursts have peak flux densities in the range $0.02$--$0.3$~Jy at $1.4$~GHz. The wide range of flux densities seen at Arecibo, some near the detection threshold, suggests that weaker bursts are also produced, likely at a higher rate \citep{spitler2016}. Although the bursts do not show any periodicity, they appear to cluster in time, with some observing sessions showing multiple bright bursts and others showing none.

The European very-long-baseline interferometry network (EVN) observations performed by \citet{marcote2017} were simultaneous with the detection of four new bursts. One of them, dubbed burst \# 2, was an order of magnitude stronger than the others. The luminosity of this burst, at the estimated distance of the host galaxy observed by \citet{tendulkar2017}, is of approximately $6\times10^{42}$ erg s$^{-1}$ at 1.7~GHz. Its associated brightness temperature is approximately $2.5 \times 10^{35}$~K, meaning more than twenty three orders of magnitude above the Compton limit, clearly indicating that the emission is coherent.

The radio observations reported by \citet{chatterjee2017} and \citet{marcote2017} show a compact source with a persistent emission of  approximately $ 180~\upmu$Jy at 1.7~GHz, implying a radio luminosity of  approximately $3 \times10^{38}$~erg~s$^{-1}$, with a bandwidth of 128~MHz. No significant, short-term changes in the flux density occur after the arrival of the bursts. Its 1--10-GHz radio spectrum is flat, with an index of $\alpha=-0.20 \pm 0.09$, $S_{\nu}\propto \nu^{\alpha}$. The projected linear diameter of the persistent radio source is measured to be less than 0.7~pc at 5.0~GHz, and it is found to be spatially coincident with the FRB~121102 location within a projected distance of 40~pc. This kind of a close proximity strongly suggests that there is a direct physical link between the bursts and the persistent source. The observed properties of the persistent source cannot be explained by a stellar or intermediate black hole, either in a binary system or not, a regular supernova remnant (SNR), or a pulsar wind nebula such as the Crab \citep[see][for a detailed discussion]{marcote2017}.

There is a 5-$\sigma$ X-ray upper limit in the 0.5-10 keV band for the radio source of $L_X\leq 5.3\times 10^{41}$~erg~s$^{-1}$ \citep{chatterjee2017}. Hence, the ratio of the radio to the X-ray flux is $\log R_X > -2.4$,  consistent with those observed in low-luminosity AGNs \citep{paragi2012}.

The host galaxy of the bursts and the persistent radio source is a dwarf star-forming galaxy with a diameter of less than 4~kpc and a high star-forming rate of $0.4\,$M$_{\odot}$~yr$^{-1}$. The stellar mass of the galaxy is estimated to be in the range $4$--$7\times10^{7}$ M$_{\odot}$. Current evidence supports the idea that the supermassive black hole-to-galaxy mass ratio lies within $0.01-0.05$ \citep{targett2012}; there are few exceptions for which this ratio can be as high as $0.15$ \citep[e.g.][]{vanderbosch2012}. Therefore, the presence of a supermassive black hole with mass $M_{\rm BH} >10^7\ $M$_{\odot}$ in the host galaxy of \frb\, is unlikely, because it would already have a mass larger than, or of the order of, the total stellar mass of the galaxy. This leads to an estimation of the mass of the putative black hole of $M_{\rm BH}\sim 10^5$--$10^6\ $M$_{\odot}$.

\section{Model and emission mechanism}\label{sect3}

A model for FRB~121102 should be capable of accounting for the fast bursts, their plurality, and the compactness of the continuum radio source. In addition, in the context of an AGN, the model should account for the modest energy budget inferred from the moderate $M_{\rm BH}$-range allowed, and the stringent X-ray upper limit. A young supernova remnant powered by a strongly magnetized and rotating neutron star faces at least three major problems: the constant DM observed in the bursts from 2012 to 2016, the lack of change of the radio flux due to rapid expansion expected in all very young SNRs, and finally the absence of an X-ray counterpart that in the case of a pulsar-powered remnant is unavoidable  \citep[see, e.g.][]{waxman2017}. We favour, instead, a model based on interactions between a relativistic (magnetized) $e^\pm$-jet launched by a  moderately massive black hole with material in the centre of the host galaxy.

\subsection{Jet-cloud  interaction}\label{jci}

We assumed that the galaxy where FRB~121102 occurs hosts a low luminosity active galactic nucleus (AGN), as suggested by \citet{chatterjee2017}. Accretion onto the central black hole results in the launching of a relativistic jet, which we assume takes place in an episodic fashion. This discontinuous jet may not be resolvable at radio frequencies; the outflow may become smooth on pc-scales, however, and be responsible for the observed persistent radio source of size $<0.7$~pc. In the jet innermost regions, on the other hand, the sporadic jet can interact with material accumulated on its way while jet activity was off. For example, at spatial scales of  approximately $10^{13}$~cm from a central black hole of $M_{\rm BH}\sim 10^5\,{\rm M_\odot}$ (see below), clouds moving at approximately  Keplerian velocities could fill the channel opened by a previous jet on day timescales.

The interaction between an episodic ejection with a cloud can take place without a cloud penetration phase into the jet. In fact, avoiding this kind of a phase is required, because it would last much longer than the actual FRB duration for any reasonable parameter choice. Another condition for the interaction is that the cloud boundaries or edges must be sharp enough for a quick jet-cloud effective interaction. In principle, the thermal or the ram pressure of the environment confining the cloud can provide this sharpening. For the same reason, the jet-leading edge should be also sharp, whereas the magnetic field should be weak enough so as to avoid rapid $e^\pm$-beam isotropization. These two conditions can naturally occur in the  scenario adopted here.

A standard ejection mechanism in AGNs is the production of magnetized jets from accreting rotating black holes \citep{blzn77}, in which the innermost regions of jets consist of strongly magnetized, relativistic $e^\pm$-beams. No significant presence of protons is expected at the base of this kind of a jet, although barions are thought to be entrained farther out \citep[ see, e.g.][and references therein]{per14}.

The different ejections of the intermittent jet should have a configuration akin to that of a relativistic ejection driven by an impulsive magnetohydrodynamic (MHD) mechanism, meaning a weakly magnetized thin leading shell moving with a high Lorentz factor driven by strong magnetic pressure gradients, followed by the strongly magnetized jet bulk \citep{komissarov2011,granot2011}. This kind of impulsive MHD acceleration mechanism allows the ejecta to achieve higher bulk Lorentz factors, $\gamma \geq 100$, than steady outflows under similar conditions. 

Because the ejection leading shell is dominated by its bulk kinetic energy, the electrons and positrons (electrons hereafter) propagate following quasi-straight trajectories as a cold $e^\pm$-beam. Thus, the electron and beam Lorentz factors can be considered equal in the laboratory frame (LF), $\gamma$. When this ultra-relativistic jet-leading edge reaches the target cloud, electrons propagate in a straight line until electric and magnetic fields cause a significant deflection. This stage, in which electrons move in quasi-straight trajectories within the cloud, presents suitable conditions for highly beamed, strong coherent emission, as long as the particles' mean free path is not too short (see below).

\subsection{Caviton formation and coherent emission}\label{cavfor}

\subsubsection{Beaming and light retardation effects}\label{blrf}

The penetration of a relativistic $e^\pm$-beam into a denser target plasma results in the formation of concentrations of electrostatic plasma waves called cavitons. The electrons crossing this caviton-filled region produce the coherent emission (see Sec. \ref{coh}). This emission is strongly beamed towards the observer if the line of sight coincides with the electron direction of motion. 

When an electron crosses a caviton, it emits a pulse of Bremsstrahlung-like radiation within a solid angle of approximately $\ 1/\gamma^2$ in its direction of motion, such as the observer direction of motion. This yields a beaming factor for the radiation of approximately $\ \gamma^2$. When particle deflection is included, however, the beaming factor towards the initial electron direction gets reduced. For instance, assuming an uniform magnetic field $B$, the average beaming factor along a distance covered by the electron equal to its gyro-radius ($r_{\rm g}=\gamma m_ec^2/qB$) yields a factor approximately $\gamma$ instead of $\gamma^2$. It should be noted that  $r_{\rm g}$ is the electron mean free path in the case of an uniform $B$-field. 

In addition to relativistic geometric beaming, light retardation effects also strongly enhance the radiation luminosity in the electron direction of motion, because the apparent time in which electrons radiate is shortened by a factor of $\ 1/2\gamma^2$ with respect to the LF. In our scenario, this factor does not need to be averaged along the electron trajectory. The reason for this is that most of the radiation towards the observer is actually produced before the electron is deflected by an angle $\gtrsim 1/\gamma$\footnote{For this very same reason, the effective beaming factor is $\sim \gamma^2$ rather than $\sim \gamma$, for source statistics purposes.}.

Both effects, the averaged beaming and the light retardation, lead to an enhancement of the apparent luminosity when looking at the beam on axis by a factor approximately $\gamma^3$ compared to the LF luminosity. It should be noted that the factor $\gamma^3$ is actually a lower limit. This is due to the assumption of an uniform $B$-field, which produces the electron strongest possible deflection. In the case of negligible electron deflection, the apparent luminosity would be enhanced by $\delta_{\rm D}^4 \approx 16\gamma^4$, where $\delta_{\rm D}$ is the Doppler factor. This is because the beaming factor would be constant along the straight electron trajectory, and of the order of  $\gamma^2$.

\subsubsection{The coherent emission mechanism}\label{coh}

Collective effects lead to coherent radiation, enhancing the emitted power \citep{weatherall1991}. For a uniform jet, the emission between any two electrons scattered by a caviton will not present any phase coherence. If, on the contrary, the jet presents density fluctuations that are correlated, then the radiation can be coherent, and therefore strongly enhanced. This correlation in the density fluctuations is the result of turbulence generated by the coupling between the background plasma and the beam: electrons from the beam perturb the plasma, producing the two-stream instability, then cavitons form, and beam-electron bunching is also generated \citep{weatherall1991}. Turbulence development should not significantly affect the cold nature of the beam as long as the turbulence-associated electron velocities do not themselves become relativistic in the flow frame. Caviton formation takes place on very short timescales,  $\ll 1$~ms in the LF \citep{beall1999}, and the temperature of the target plasma is expected to rise similarly fast. Thus the mechanism could work even for the very short timescales of FRBs.

\subsubsection{Electron deflection and flow isotropization}\label{def}

The residual magnetic field dragged along by the jet-leading thin shell eventually deflects the energetic electrons propagating through the target plasma. The interpenetration of the jet-leading thin shell with the target cloud leads to the formation of a contact layer between both, the region in which the coherent emission is produced, in which electrons tend to isotropize. This layer is approximately at rest in the LF, and a strong enhancement of the perpendicular $B$-component ($B_\perp$) dragged by the beam is expected there. For this reason, it seems natural to assume that the mean free path of the electrons in the interpenetration region will be approximately $r_{\rm g}$, with $B\sim B_\perp$. On the other hand, the electrostatic field $E_0$ inside the cavitons, expected to be randomly oriented, will induce pitch-angle diffusion to the emitting electrons. The value of $E_0$ cannot be too high, because the mean free path of electrons should be larger than the caviton size, meaning $r_{\rm g}>D$, if the mechanism is to work. More precisely, $r_{\rm g}\gg D$ if pitch-angle diffusion is realized. For the same reason, the magnetic field in the beam cannot be too high, and a similar constraint applies to the cloud magnetic field, the geometry of which may be between a strongly ordered $B$-field and the chaotically oriented $E_0$-field. It is also worth noting that, even if the electron mean free path in the interpenetration region is $>D$, it should also be  long enough as to ensure that the emitting volume is sufficiently large to produce the observed fluxes.

The time required for particle isotropization in the beam is $\delta t_{\rm iso}\gtrsim r_{\rm g}/c$. After this time, the perturbation originated by the jet-cloud interaction can affect the incoming electrons even before it reaches the cloud, isotropizing them and the electromagnetic fields they carry. At this stage, the isotropized electrons may still interact with cavitons around the jet-cloud contact discontinuity through some level of jet-cloud interpenetration. However, the related coherent emissivity is strongly reduced by a factor of $\gtrsim 1/\gamma^3$ (see Sect.~\ref{blrf}). Moreover, as the bulk of the jet, which is more magnetized than its leading edge, reaches the cloud, the expected higher magnetic field filling the region can stop the coherent radiation completely by making the electron mean free paths smaller than the cavitons. We conclude then that $\delta t_{\rm iso}$ determines the duration of the coherent emission phase for electrons from the jet-leading thin shell interacting simultaneously with the cloud, such that $\delta t_{\rm obs}\sim 1\,{\rm ms}\gtrsim\delta t_{\rm iso}$.

\subsubsection{Event duration}

Despite $\delta t_{\rm iso}$ playing a major role in the FRB scenario proposed here, it seems likely that the cloud will present an irregular surface. Assuming that $\Delta r$ is the relevant irregularity scale of the target cloud, the ultra-relativistic thin shell must cover a distance $ \Delta r$ in a time $\delta t_{\rm cross}\approx \Delta r/c$ in the LF for full interaction. Unless $B$ is extremely low, it will be the case in our scenario that $\delta t_{\rm iso}\ll 1$~ms, in which case $\Delta r$ could determine the event duration. This implies that, for the observer, $\delta t_{\rm obs}\approx \delta t_{\rm cross}/2\gamma^2\approx \Delta r/2c\gamma^2\lesssim r_{\rm j}/2c\gamma^2$, where $\Delta r\approx r_{\rm j}$ is the largest effective irregularity scale, and light retardation effects have been taken into account, meaning: $\delta t_{\rm obs}\approx \delta t_{\rm cross}/2\gamma^2$. 
Therefore, unless the magnetic field is very low, it is the case that $\delta t_{\rm iso}/c\ll 1$~ms, and the duration of the burst will be determined by the crossing time of the irregularity scale corrected by light propagation effects. The dynamical timescale of the intermittent ejections does not affect the duration of the burst if it is $\gtrsim 1$~ms, which is expected given that the light-crossing time of the central black hole is approximately $1\,(M_{\rm BH}/10^5\,M_\odot)$~s. Even if the jet ejection lasts for $\gg 1$~ms, the isotropization of the beam particles will stop, or at least strongly reduce, the coherent emission.

\subsection{Radiation properties}

The spectrum of the coherent emission presents two main components, a line at the plasma frequency $\omega_e$,  and a broad-band tail that inherits the power-law behaviour of the density fluctuation spectrum of the $e^\pm$-beam. The existence of this type of emission is well known from controlled plasma experiments \citep[e.g.][]{kato1983,masuzaki1991,benford1992,ando1999}.

The broad-band component of the spectrum extends from the plasma frequency  to $\omega_{\rm{max}} = 2 \gamma^2 c/D$, which is the highest frequency emitted by electrons with Lorentz factor $\gamma$ being scattered by cavitons of size $D$. The size $D$ of the  cavitons induced in the plasma by the relativistic $e^\pm$-beam can be estimated as approximately $ 15\,\lambda_{\rm D}$ \citep{weatherall1991}, where  ${\lambda_{\rm D}}$ is the Debye length of the plasma, such that $\lambda_{\rm D}= 6.9\sqrt{T_{\rm c}/n_{\rm c}}$~cm, with $[T_{\rm c}] =$ K and $[n_{\rm c}] =$ cm$^{-3}$.

For a $e^\pm$-beam interacting with a denser target plasma, the required condition for collective emission is $q = n_{\rm b}/n_{\rm c} \ge 0.01$, with $n_{\rm b}$ and $n_{\rm c}$ being the beam and target densities, respectively. The radiated power per electron in the LF in the form of coherent emission is given by \citep{weatherall1991}:
\begin{equation}\label{eq:power}
        P_e =  \frac{E_0^2 \sigma_{\rm{T}} c}{8 \pi}  \frac{4 n_{\rm{b}} \pi D^3}{3}  \frac{27 \pi}{4} f
        \left[1 + \left( \frac{\Delta n_{\rm{b}}}{n_{\rm{b}}} \right)^2 0.24 \ln \left( \frac{2 \gamma^2 c}{D \omega_p} \right)\right],      
\end{equation}
where $\sigma_{\rm{T}}$ is the Thomson cross-section, $\Delta n_{\rm{b}}/n_{\rm{b}}$ is the fluctuation-to-mean-density ratio of the relativistic electrons, and $f$ is the fraction of the cloud volume that is filled with cavitons, for which we adopted $f=0.1$ \citep{levron88}. The second factor in Eq.~(\ref{eq:power}), $4 n_{\rm{b}} \pi D^3/3$, is the number of electrons inside a caviton, and it is known as the coherence factor \citep{weatherall1991}.

The first term in Eq.~(\ref{eq:power}) represents the power emitted at the plasma frequency, whereas the second term is the power emitted in the tail of the spectrum. Emission at the plasma frequency is likely to be absorbed \citep{weatherall1991}, hence the relevant term is the one associated with the broad-band emission. It is natural for a  density fluctuation spectrum with power-law behaviour to arise as the result of turbulence in the plasma, for example a Kolmogorov spectrum has an index $\alpha = -5/3$. Density fluctuations show a large range of variation, and can reach values of order unity, meaning  $\Delta n_{\rm{b}}/n_{\rm{b}} \sim 1$ \citep{henri2011}.  For these kinds of density fluctuations, the radiation spectrum behaves as $\nu P_{\nu}\propto \nu^{\alpha+1}$, and the power emitted in the broad-band component is comparable to the power emitted at the line, with most of the former being radiated in the low-frequency part of the power-law spectrum. Thus, although the radiation produced at the plasma frequency may be easily absorbed by the emitting plasma, the radiation at slightly higher frequencies has a comparable luminosity. The total power $P_{\rm t}$ is simply the power emitted per particle times the number of relativistic electrons inside the region filled with cavitons.

\section{Application to FRB~121102}\label{sect4}

\begin{figure}[ht]

\begin{center}

\includegraphics[width=0.9\linewidth]{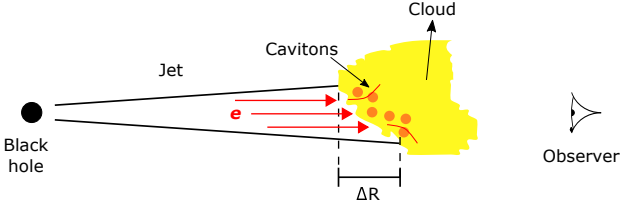}

\caption{Schematic representation of the model. }

\label{fig:jet}

\end{center}

\end{figure}

In the following, we consider a cloud with an irregular surface with irregularity scale $\Delta r\approx r_{\rm j}$ and total jet-interaction section $\pi\,r_{\rm j}^2$ (see Fig.~\ref{fig:jet}). As explained in Sect.~\ref{cavfor}, $\Delta r$ and, consequently, $r_{\rm j}$ are constrained by the duration of the event:
\begin{equation}\label{eq:rj}
\Delta r \approx r_{\rm j}\approx 6 \times 10^{11} \Big( \frac{\gamma}{100} \Big)^2\,\rm{cm}\,.
\end{equation} 
Assuming a half-opening angle of a few degrees, this kind of an $r_{\rm j}$-value would correspond to approximately $10^3\,R_{\rm G}\sim 10^{13}$~cm for a $10^5\,{\rm M_\odot}$ black hole, where $R_{\rm G}=GM/c^2$.

The coherent emission is broad-band, extending from the plasma frequency $\omega_e$ up to $\omega_{\rm{max}} = 2 \gamma^2 c/D$. The observed frequency of $1.7$ GHz should be within the emission range $ \omega_e/2\pi < 1.7 \textrm{ GHz} <  \omega_{\rm{max}}/2\pi$. The first condition, $ \omega_e /2\pi < 1.7$ GHz, results in $n_{\rm c} \leq 4 \times 10^{10}$ cm$^{-3}$, where the subscript c indicates the region where cavitons form.

Because the emission radiated at the plasma frequency will likely be absorbed, we adopted a lower value of the plasma density, $n_{\rm c} = 10^{10}$~cm$^{-3}$. With this choice, the peak of the emission is at $\nu = 900$ MHz. If one adopts a Kolmogorov spectrum for the turbulence, $\nu P_{\nu} \propto \nu^{-2/3}$, the luminosity at the observed frequency will be approximately $70$\% of the luminosity of the peak.

It is worth mentioning that calculations in the weak turbulence regime suggest that ambient plasma might also, in principle, produce an attenuation of the coherent signal by Raman scattering \citep[e.g.][]{levinson95}. However, experiments show that these effects are suppressed in the strong turbulence case, where there is no theory available \citep{benford1998,romero2016}.

For the coherent emission to escape, free-free absorption should be also minor within the cloud. Adopting $6\times 10^{11}$~cm for the cloud size and $n_{\rm c}= 10^{10}$~cm$^{-3}$, the cloud is optically thin as long as its (pre-interaction) temperature is $\gtrsim 10^8$~K; in principle this is possible because the virial temperature at $10^3\,R_{\rm G}$ is approximately $10^{10}$~K. 

The second condition, $1.7  \textrm{ GHz} < \omega_{\rm{max}} /2\pi$, results in:
\begin{equation}\label{eq:second}
        \gamma^2 \sqrt{ T_{\rm c} /n_{\rm c}} \geq 73.7\,,
\end{equation}
with $[T_{\rm c}] =$ K and $[n_{\rm c}] =$ cm$^{-3}$. The impact of the jet can heat the cloud up to a temperature of $T_{\rm c} = q m_e c^2 \gamma/k_{\rm B}$. For the adopted $n_{\rm c}$-value, and a density ratio of $q = 1$, $ T_{\rm c} = 6 \times 10^{11}$~K; this $T_{\rm c}$-value together with the adopted one for $n_{\rm c}$ fulfils Eq.~(\ref{eq:second}). Cavitons formed in a plasma with these characteristics have a size of $D \sim 800$ cm; the number of electrons inside cavitons results in $2 \times 10^{19}$. The obtained value for $D$ is similar to the values found in particle-in-cell (PIC) numerical simulations \citep[e.g.][]{beall2010}. It is worth recalling that coherent emission is only possible if $r_{\rm g}$ is larger than the size of cavitons, as discussed in Sect. \ref{def}.

The jet power can be obtained from the jet particle density in the LF at the interaction location, $n_{\rm j}$, through the relation:
\begin{equation}\label{eq:n}
        \gamma n_{\rm j} m_e c^2  \approx \frac{L_{\rm j}}{\pi r_{\rm j}^2 c}\,.
\end{equation}
Using the radius jet constrained by the event duration given by Eq. \ref{eq:rj}, Eq.~(\ref{eq:n}) yields a jet power $L_{\rm j}\sim 3 \times 10^{40}$~erg~s$^{-1}$.

When electrons enter the cloud, they strongly radiate towards the observer until they isotropize, which occurs on a time $\delta t_{\rm iso} \gtrsim r_{\rm g}/c\sim 6\times 10^{-8}\,\gamma/B$~s. Assuming that the magnetic energy density is a fraction $\xi$ of the jet kinetic energy density, the magnetic field in the LF can be obtained from:
\begin{equation}\label{eq:B}
        \frac{B^2_{\rm j,eq}}{8 \pi} = \frac{1}{2}\frac{\xi L_{\rm j}}{\pi r_{\rm j}^2 }\,.
\end{equation}
At the interaction location, this yields an equipartition field (i.e. $\xi =1$) of $B_{\rm j,eq}\sim 3200$~G. We adopted a magnetic field well below equipartition ($\xi  < 1$) in the interacting shell, as explained in Sect.~\ref{jci}; in particular, we consider $B = 5 \times 10^{-2}B_{\rm j,eq}$.

To obtain $E_0$ in Eq.~(\ref{eq:power}), we imposed the condition that the pitch-angle diffusion timescale, given by
\begin{equation}\label{eq:diffE}
        t_{\rm diff} = \frac{\lambda_{E_0}^2}{D c},
\end{equation}
with $\lambda_{E_0} = \gamma m_e c^2/eE_0$, should be longer than $\delta t_{\rm iso} = r_{\rm g}/c$, as discussed in Sect. \ref{def}). This results in $E_0 \sim 45$ G, yielding $W = E_0^2/8 \pi n_{\rm c} k_{\rm B}T_{\rm{c}} \sim 10^{-4}$, which is within the range values obtained in numerical simulations \citep{henri2011}.

Typically, $\delta t_{\rm iso}<< \delta t_{\rm cross}$, and therefore not all electrons will radiate simultaneously. By the time the jet-leading edge has crossed all of the cloud irregular surface, the electrons that interacted first are already isotropized and their coherent emission has been suppressed. To derive the observer luminosity, the LF luminosity (Eq.~\ref{eq:power}) must be corrected by a factor of $\delta t_{\rm iso}/\delta t_{\rm cross}$ to account for a smaller simultaneously emitting volume. In addition, to account for light retardation effects, another factor $t_{\rm cross}/t_{\rm obs}$ must be considered. All this renders a factor  of$\delta t_{\rm iso}/\delta t_{\rm obs}$ that for the adopted values of $\Delta r$ and $B$ is $\gtrsim 3\times 10^{-5}$. In addition to  $t_{\rm iso}/t_{\rm obs}$, Doppler boosting must be considered to obtain the observer luminosity, as explained in Sect.~\ref{cavfor}; this results in:
\begin{equation}
L_{\rm obs}\approx \gamma \Big(P_e N_e \Big) \frac{\delta t_{\rm iso}}{\delta t_{\rm obs}}\,,
\end{equation} 
where $N_e$ is the number of electrons in the volume of the emitting region, meaning approximately $\pi\,r_{\rm j}^2\,r_{\rm g}$.

Using the adopted parameter values, the predicted FRB observer luminosity is approximately $3 \times 10^{42}$~erg~s$^{-1}$, comparable to the luminosity of burst \#2 of FRB~121102. The values of the parameters adopted above are just one of several possible choices, because different combinations of plausible values can also explain burst \#2 and other bursts with different properties.

We note that for extremely low $B$-values, $\Delta r<r_{\rm g}$, in which case $\delta t_{\rm iso}$ and not $\delta t_{\rm cross}$ would determine $\delta t_{\rm obs}$. This would correspond to a plane jet-cloud interface, in which case no weighting by $\frac{\delta t_{\rm iso}}{\delta t_{\rm obs}}$ should be applied.

Another possibility that cannot be ruled out is that the transition region between the unisotropized, meaning unshocked, and the isotropized, meaning shocked, beam could be the coherent emitter  \citep{weatherall1991}. However, the magnetic field in that location might be too strong if the bulk of the jet is highly magnetized. In addition, the duration of the FRB in this scenario would have to be determined by some ad hoc mechanism.

\subsection{Alternative scenario: a lighthouse effect}

An alternative to the scenario presented here, in which the FRBs occur at the onset of a jet's ejections, is that of a jet changing its direction \citep[see][]{katz2016b}. Occasionally, this wandering jet  would intercept a cloud\footnote{In the model discussed in \citet{katz2016b} the jet does not necessarily intercept a cloud; in this case, the jet might sweep through the existing medium or the radiation might be produced internally in the jet.} while  pointing to the observed and then producing an FRB. In that case, the timescale of the event will be the time needed by the jet to change direction by $1/ \gamma$~rad while it is interacting with the cloud. This situation resembles that of a lighthouse, in which the event duration is not affected by causality constraints.

A lighthouse effect combined with coherent emission from jet-target interactions cannot in principle be discarded, and dispenses us with the need to assume a discontinuous jet with a sharp leading edge interacting with an irregular target. On the other hand, a modest magnetic field and a high Lorentz factor are still required. In fact, most of the details of the model given in Sect. \ref{sect3} still hold in this alternative scenario, but now the crossing and the observer timescales are the same. 

Nevertheless, there is drawback in the lighthouse scenario. The change in direction by $1/ \gamma$~rad of the emitting electrons in just $1$~ms requires that the properties of the leading thin shell substantially change along the jet axis on a spatial scale  of approximately $1\,{\rm ms}\cdot c\sim 3\times 10^7$~cm, which is $\ll r_{\rm j}$. This kind of a jet configuration is in principle possible but requires the flow to be very cold, meaning a Mach number $\gg 1/\gamma$. Otherwise, electrons will tend to homogenize their properties on these small scales, and the very short-scale jet-bending coherence will be lost. In addition, the fast changes in direction require angular velocities of approximately $10\,(100/ \gamma)$~rad~s$^{-1}$, which may not be feasible for a jet-launching engine that has already a size $\gtrsim 1\,(M_{\rm BH}/10^5\,M_\odot)$~light-second. 

The constraints mentioned can be relaxed for smaller central engines and thus smaller black hole masses, although smaller black holes imply tighter energy limits. If the source were super-Eddington, for example similar to SS~433, the energetics might fulfil the minimum requirements \citep{katz1980,katz2016b,kaufman2002}, but this kind of a scenario requires a dedicated study.

\section{Discussion}\label{sect.5}

We propose that FRB 121102 and similar events are the result of coherent radio emission produced by a relativistic, turbulent $e^\pm$-beam interacting with plasma cavitons. An advantage of this mechanism is that it might operate in different scenarios involving relativistic jets. For instance,  \citet{romero2016} discuss possible settings involving long gamma-ray bursts and mini-jets produced by magnetic reconnection inside a larger outflow. Even single-event FRB may be explained in the basic framework of the proposed model as long as the recurrence time of the events is very long, depending on sensible factors such as beam orientation, Lorentz factor, propagation length within the plasma, and the interaction scale. Here, we investigate the mechanism in the context of an extragalactic episodic jet interacting with the environment, to check whether the model can account for FRB~121102 in light of  new observational evidence. In what follows we further comment on a number of important assumptions of the model.

\subsection{Sporadic ejections}

The proposed scenario requires episodic jet launching. Episodic ejections are known to take place in several astrophysical sources. The hydrogen ionization instability is responsible for switches between periods of outburst and quiescence in dwarf novae. The state transition observed in numerous X-ray binaries is also proof of a variable accretion regime \citep[e.g.][]{done2007}. In fact, multiple variability timescales are common for the radiation associated with galactic and extragalactic jets. Non-steady jet production may be behind this variability and therefore render it a somewhat natural phenomenon, at least at the relatively small spatial scales relevant in our scenario, meaning approximately $10^3\,R_{\rm G}$. At larger scales this sporadic jet activity does not affect the persistent radio source.

\subsection{High jet Lorentz factors}

As indicated, strongly magnetized sporadic ejections can be efficiently accelerated by a MHD-driven impulsive mechanism to high bulk Lorentz factors such as $\gamma \geq 100$ \citep{granot2011}. Non-stationary magnetized outflows have also been proposed to explain the apparent disagreement between the typical AGN Lorentz factors inferred from radio data, $5 \lesssim  \gamma \lesssim 40$ \citep{jorstad2005}, and the higher Lorentz factors invoked to explain the rapid TeV-variability observed in blazars \citep[e.g.][]{barkov2012}. In this context, the short timescale flares observed at TeV energies would be associated with variable emission from these shells, whereas the radio data would be associated with the emission from a larger scale, smoother flow with a lower Lorentz factor \citep{lyutikov2010,komissarov2011}.

\subsection{Cloud origin}

Clouds from the AGN broad-line region (BLR) present densities of $10^{10}-10^{11}$~cm$^{-3}$ at distances of $10^3-10^4\,R_{\rm G}$ to the central black hole \citep{peterson2006,risaliti2011}.  The presence of material for jet interaction might be also related to the accretion phenomenon itself \citep[e.g.][]{bla99,beg12}.

\subsection{Other observational aspects}

\subsubsection{The persistent radio source}

In our model, the observed continuum flat-spectrum radio source would correspond to the synchrotron radiation of  the  jet, which is the result of the averaged intermittent ejections. The synchrotron luminosity expected at radio wavelengths  for a jet with $L_{\rm j}=3 \times 10^{40}$~erg~s$^{-1}$ can be roughly estimated as \citep[e.g.][]{bosch-ramon2015}:

\begin{equation}
L(\sim 1.7 \textrm{ GHz})\approx \eta_{\rm NT}L_{\rm j}\frac{\delta_{\rm D}^4}{\gamma^2}\frac{t_{\rm esc}}{t_{\rm syn}},
\end{equation}

\noindent where $\eta_{\rm NT}$ is the non-thermal-to-total energy density ratio in the emitter, $\delta_{\rm D}$ is the Doppler factor, and $t_{\rm esc}$ and $t_{\rm syn}$ are the electron escape and cooling times, respectively. As discussed in Sect. \ref{blrf}, for a highly relativistic jet pointing towards the observer, $\delta_{\rm D} \sim 2\gamma$. The jet magnetic field can be estimated assuming again a certain value for the equipartition fraction $\xi$, not necessarily the same as in the FRB-emitting region. From all this, plus adopting a jet distance to the black hole of $z \sim 1$~pc and a jet half-opening angle of $1/\gamma=0.1\,(10/\gamma)$~rad, one obtains:

\begin{equation}
L(\sim 1.7 \textrm{ GHz})\approx 6 \times 10^{39} \textrm{erg s}^{-1} \Big(\frac{\eta_{\rm NT}}{10^{-1}}\Big) \Big(\frac{\gamma}{10} \Big)^2 \Big(\frac{\xi}{10^{-1}} \Big)^{3/4}.
\end{equation}

This luminosity is well above the persistent radio luminosity mentioned in Sect.~\ref{sect2}, indicating that the scenario under typical assumptions is consistent from the energetic point of view even when considering duty cycles of jet activity $\lesssim 10$\%. It is worth noting that the jet luminosity obtained in Sect.~\ref{sect4} is also compatible with the X-ray upper limit.

\subsubsection{Black hole mass}

The stellar mass of the galaxy is in the range $4$--$7\times10^{7}$ M$_{\odot}$ \citep{tendulkar2017}. Little is known about the existence of massive black holes in these kinds of small galaxies. If we extrapolate from the scaling relation given by \citet{reines2015}, determined in the range $10^8 \leq M_{\rm stellar}/{\rm M_{\odot}} \leq 10^{12}$, we obtain a black hole mass of approximately  $ 10^5$ M$_{\odot}$, within the allowed range (see Sect.~\ref{sect2}). Although most of the estimations of the masses for black holes in the centre of galaxies are above $10^6~{\rm M_{\odot}}$, there is evidence of the presence of black holes with masses of  approximately $10^5$ M$_{\odot}$ in some AGNs \citep{papadakis2004}. If this kind of a black hole accretes at 1\% of the Eddington rate, its luminosity would  be approximately $10^{41}$~erg~s$^{-1}$, a value comparable to the one adopted in our model, and of the order of the X-ray upper limit.

\subsubsection{Polarization}

FRBs present an additional challenge concerning polarization. There is no evidence of polarized emission from the repeater \frb ~\citep{scholz2016}; however, FRB~150807 presented linear polarization \citep{ravi2014}, whereas a high degree of circular polarization was measured in  FRB~140514 \citep{petroff2015}. The radiation mechanism proposed in this work might produce linear polarization in the presence of a magnetic field, whereas intrinsic circular polarization is not expected \citep{benford1992b}.

\subsubsection{High-energy emission}

An analysis of the multi-wavelength non-thermal emission associated with the mechanism discussed for FRB~121102 is under way and will be presented elsewhere. We can put forward however a general framework in which the cloud impacted by the jet and the shock in the jet itself may lead to efficient particle acceleration, and to high-energy emission that could be detectable very briefly, seconds to minutes, if the beam is fast enough, even for modest energetic budgets  \citep[see, e.g.][]{barkov2012,bosch-ramon2015}.

\section{Conclusions}\label{sect.6}

The model proposed in this work, based on a mechanism of coherent emission in beam-excited plasma cavitons, is able to explain the diverse properties of FRBs: the extragalactic origin, the energy budget, the high brightness temperature, the repetitions with no apparent periodicity, and the counterparts and upper flux limits obtained in different wavelengths. The very short duration of the events is explained by the dynamical timescale of the process corrected by light retardation effects, although the isotropization timescale of the beam particles plays also an important role, and may determine the event duration for very low magnetic fields.

\section*{Acknowledgments}

The authors are grateful to Jordi Miralda-Escud\'e for his very useful comments and suggestions. This work was supported by the Argentine Agency CONICET (PIP 2014-00338) and the Spanish Ministerio de Econom\'{i}a y Competitividad (MINECO/FEDER, UE) under grants AYA2013-47447-C3-1-P and AYA2016-76012-C3-1-P with partial support by the European Regional Development Fund (ERDF/FEDER), MDM-2014-0369 of ICCUB (Unidad de Excelencia `Mar\'{i}a de Maeztu'), and the Catalan DEC grant 2014 SGR 86. V.B.R. also acknowledges financial support from MINECO and European Social Funds through a Ram\'on y Cajal fellowship. This research has been supported by the Marie Curie Career Integration Grant 321520. M.V.d.V acknowledges support from the Alexander von Humboldt Foundation.

\bibliographystyle{aa}

\bibliography{myrefs8}   

\end{document}